\documentclass[12pt,preprint]{aastex}
\usepackage{psfig}

\newcommand\mdot   {\hbox {${\dot M}$}}

\newcommand\pp     {$\pm$}
\newcommand\pers   {s$^{-1}$}
\newcommand\micros {$\mu$s}

\def\degr{\hbox{$^\circ$}}

\newcommand\funit   {erg cm$^{-2}$ s$^{-1}$}

\righthead{State transitions in SAX J1711.6--3808}
\slugcomment{Accepted by ApJ Main Journal: 13 September, 2001}

\begin{document}

\title{The new X-ray transient SAX J1711.6--3808: decoupling between
its 3--20 keV luminosity and its state transitions}

\author{Rudy Wijnands\altaffilmark{1,2} \& Jon M. Miller\altaffilmark{1}}

\altaffiltext{1}{Center for Space Research, Massachusetts Institute of
Technology, 77 Massachusetts Avenue, Cambridge, MA 02139-4307, USA;
rudy@space.mit.edu; jmm@space.mit.edu}

\altaffiltext{2}{Chandra Fellow}

\begin{abstract}

We present a study of the correlated spectral and timing behavior of
the new X-ray transient SAX J1711.6--3808 during its 2001 outburst
using data obtained with the {\it Rossi X-ray Timing Explorer} ({\it
RXTE}). We also investigate the correlations between those source
properties and the 3--20 keV X-ray luminosity.  The behavior of the
source during the observations can be divided into two distinct state
types. During the ``hard'' state, the energy spectra are relatively
hard and can be described by only a power-law component, and the
characteristic frequencies (i.e., the frequency of the 1--7
Hz quasi-periodic oscillations observed for the first time in this
source) in the power spectra are low. However, during the ``soft''
state, the spectra are considerably softer (in addition to the
power-law component, a soft component below 8 keV is necessary to fit
the spectra) and the frequencies are the highest observed. Remarkably,
this distinction into two separate states cannot be extrapolated to
also include the 3--20 keV X-ray luminosity. Except for one
observation, this luminosity steadily decreased but the hard state was
observed both at the highest {\it and} lowest observed
luminosities. In contrast, the soft state occurred only at
intermediate luminosities.  This clearly demonstrates that the state
behavior of SAX J1711.6--3808 is decoupled from its X-ray luminosity
and that if the X-ray luminosity traces the mass accretion rate in SAX
J1711.6--3808, then the state transitions are not good accretion rate
indicators, or vice versa.  The {\it RXTE} data of SAX J1711.6--3808
does not allow us to conclusively determine the exact nature of the
compact object in this system. The source resembles both neutron star
and black hole systems when they have low luminosities. We discuss our
results with respect to the correlated timing and spectral behavior
observed in other low-mass X-ray binaries and the implications of our
results on the modeling of the outburst light curves of X-ray
transients.

\end{abstract}


\section{Introduction \label{intro}}

From simple accretion theory, one would expect that the X-ray
luminosity in low-mass X-ray binaries (LMXBs) is directly related to
the mass accretion rate\footnote{When we discuss the mass accretion
rate we implicitly assume the mass accretion rate through the disk and
not through, e.g., radial inflow} (\mdot) in those systems. The X-ray
spectra and the rapid X-ray variability of LMXBs most likely originate
very close to the compact objects in those systems, in an environment
which is dominated by changes in the mass accretion rate. Therefore,
from this simple picture, the X-ray luminosity of LMXBs, their
spectra, and their rapid variability should be directly related to
\mdot~and, thus, to each other.  However, strong evidence is available
that this scenario is an oversimplification of the behavior observed
in LMXBs.

Especially for the persistent neutron-star LMXBs, it is well known
that their timing behavior is very well correlated with their spectral
behavior (as determined from the behavior in X-ray color-color
diagrams), but much less well with the X-ray luminosity. This is also
true for the transients systems for which enough data have been
obtained to correlate the different parameters. For the neutron star
systems, it is thought that the spectral and timing behavior are very
well correlated with \mdot~and that the X-ray luminosity is not. Due
to an (as yet) unidentified process (or processes), a considerable
fraction of the implied X-ray flux does not reach us (see van der Klis
1995, 2000 for detailed overviews and discussions about the neutron
star LMXB behavior).

Until recently, the above described simple picture was thought to be
an accurate description of the behavior of black-hole candidate (BHC)
LMXBs, which could be described using the concept of source states
(Tanaka \& Lewin 1995; van der Klis 1995).  In the BHC low-state (LS),
\mdot~is low, the energy spectra are hard, and the power spectra are
dominated by a very strong (20\%--50\% rms amplitude) band-limited
noise.  In the high state (HS), \mdot~is higher, the spectra are much
softer, and in the power spectra only a weak (a few percent) power-law
noise component is present. In the very high state (VHS), \mdot~is the
highest, the spectra are harder but not as hard as in the LS, and in
the power spectra, noise is present similar to either the weak HS
noise or the LS band-limited noise (although only with a strength of
1\%--15\% rms).  Quasi-periodic oscillations (QPOs) near 6 Hz are
detected, sometimes with a complex harmonic structure.

However, it has become clear that the behavior of BHCs is far more
complex than previously thought. First of all, QPOs are now observed
not only during the hard VHS but also during hard states when the
source luminosity is much less than during the VHS (e.g. the so-called
intermediate states; M\'endez \& van der Klis 1997; M\'endez, Belloni,
\& van der Klis 1998; Remillard et al. 1999b; Sobczak et al. 2000;
Dieters et al. 2000; Homan et al. 2001a), but also a different type of
QPO has been observed with frequencies up to 450 Hz (e.g., Remillard
et al. 1999a,b; Homan, Wijnands, \& van der Klis 2000; Cui et
al. 2000; Homan et al. 2001a; Strohmayer 2001; Miller et
al. 2001). However, the very detailed study of the new X-ray transient
XTE J1550--564 performed by Homan et al. (2001a), using an extensive
data set obtained with the proportional counter array (PCA) on board
the {\it Rossi X-ray Timing Explorer} ({\it RXTE}) demonstrates best
our lack of understanding of the behavior in BHCs. They showed that at
least in this source (and possibly in more BHCs), the state behavior
is completely decoupled from the X-ray luminosity and the state
behavior of XTE J1550--564 must be described using two-dimensional
diagrams. Besides the mass accretion rate, an unknown second parameter
(e.g., the inner disk radius, the size of the Comptonizing region, the
accretion flow geometry; Homan et al. 2001a) is involved in the
formation of the states.

In this {\it Letter}, we discuss the correlated spectral and timing
behavior of the new X-ray transient SAX J1711.6--3808 (in 't Zand,
Kaptein, \& Heise 2001) as observed with the {\it RXTE}/PCA, and the
lack of correlation between its state transitions and the X-ray
luminosity.

\section{Observations and analysis}

The new X-ray transient SAX J1711.6--3808 was discovered in February
2001 with the Wide Field Camera unit 1 on board {\it BeppoSAX} (in 't
Zand et al. 2001). The source flux was between 30 and 80 mCrab (2--9
keV). The {\it RXTE} all sky monitor (ASM) light curve (not shown, but
can be obtained from http://xte.mit.edu/ASM\_lc.html) shows that the
peak ASM count rate was between 4 and 5 counts \pers, which
corresponds to 50--70 mCrab (1.5--12 keV). Therefore, this source can
be classified as a weak X-ray transient. During the first part of its
outburst, proprietary {\it RXTE}/PCA observations were performed and
are not yet publicly available. However, during the latter part of the
outburst (i.e., the decay), several public TOO observations were
performed which we used in our analysis (see Tab.~\ref{tab:log} for a
log of the observations). During these observations, data were taken
simultaneously in the Standard 1 (1 energy channel and 1/8 s time
resolution) and 2 modes (129 channels; 16 s resolution), and the event
mode E\_125US\_64M\_0\_1S (64 channels; 122\micros~resolution). We
used the Standard 2 data to create a 2.9--18.8 keV count rate curve of
the source and to create color curves. We used as soft color the count
rate ratio between 4.1--7.5 keV and 2.9--4.1 keV, and as hard color
the one between 11.4--18.8 keV and 7.5--11.4 keV. 

Due to the different number of detectors (proportional counter units
or PCUs) on during our observations, we only used data of the PCU 2,
which was always on during our observations. PCU 0 was also
continuously on during our observations, however this detector has
lost its propane volume, which results in a higher background and
occasional background flares. For weak sources like SAX J1711.6--3808,
this would have considerable impact on the spectral results and,
therefore, we have not included the data of this detector in our
analysis. Although including only PCU 2 in our analysis gives limited
results on an observation-by-observation basis, comparing the spectral
results obtained with differing numbers of PCUs might introduce
unwanted systematical effects. To obtain an homogeneous set of
results, without extra systematic errors, we only use the data of PCU
2.

Routines from release 5.0.4 of the FTOOLS package were used to extract
and reduce Standard 2 spectra from PCU 2 (data from all PCU layers
were considered simultaneously).  The tool ``pcabackest'' was used to
extract background spectra, using the ``bright source'' model.
Response matrices were generated using the tool ``pcarsp.''  Fits to
PCA spectra of the Crab (a simple power-law at PCA resolution) often
show significant residuals below 3.0~keV; for this reason we set
3.0~keV as a lower fitting limit.  The source spectra become dominated
by the background above 20~keV, so we adopt 20.0~keV as the upper
limit to our fitting range.  Power-law fits to the PCA spectra of the
Crab show energy dependent residuals which can be as large as 1.0\%.
We therefore add 1.0\% systematic errors to the spectra before
fitting.  The background-subtracted 3.0--20.0~keV X-ray spectra were
analyzed using XSPEC version 11.01 (Arnaud \& Dorman 2000).  Errors on
the fit parameters were calculated for the 90\% confidence limits.
The fluxes were obtained by measuring the flux at the best-fit values
and the flux errors by measuring the flux at the 90\% confidence
limits for the given parameters (note that this method for obtaining
flux errors will give the largest errors for those data sets which
needed the most fit components; i.e. the third and fourth
observations).

The absorption column was fixed to the Galactic column density towards
this source (the weighted average obtained with the web-based tool nH
is $1.4\times10^{22}~{\rm cm}^{-2}$; Dickey \& Lockman
1990\footnote{nH can be found at
http://heasarc.gsfc.nasa.gov/cgi-bin/Tools/w3nh/w3nh.pl; after
submission of our paper, we learned that via an {\it XMM-Newton}
observation of SAX J1711.6--3808 a column density was found which is
consistent with what we used in our analysis (M. Santos-Lleo 2001
private communication)}).  However, for observation 60407-01-07-00
unacceptable fits were obtained when fixing the column density to this
value.  For this observation, the column density was a free parameter
in the fits and we obtained a column of $4.2\pm0.5\times10^{22}~{\rm
cm}^{-2}$.  The inclusion of extra components in fits (e.g., a soft
component) to observation 60407-01-07-00 using the column density
assumed for other observations did not allow for statistically
acceptable fits.  Similarly, using the column density obtained for
observation 60407-01-07-00 in fits to the other observations did not
yield significant statistical improvements, or allow the parameters of
a soft flux component to be constrained in observations wherein a soft
flux component is not indicated using the assumed column density (see
below).

The X-ray spectra of LMXBs can be fit with a variety of models (e.g.,
Christian \& Swank 1997). However, SAX J1711.6--3808 is a potential
black-hole candidate and the X-ray spectra of those sources are
usually fitted with a two-component model: a multi-color disk (MCD)
black-body and a power-law. A similar amount of consensus is not
present for the fit models for neutron star systems. Therefore, we did
not concern ourself with other spectral models but only use the MCD
plus power-law model for SAX J1711.6--3808. If this source contains a
neutron star we might obtain unusual results via this model. By
fitting the spectra with this model, a clear excess was apparent
around 6--7 keV, which we fitted with a Gaussian with fixed centroid
energy (6.4 keV) and width (1.2 keV). Although the equivalent width of
this feature could be quite large ($>$0.5 keV), it is unclear what
contribution is due to remaining calibration uncertainties of PCU 2
and/or due to the fit model assumed (note, that the strength and the
variability of this feature might indicate that at least part of it is
not due to remaining calibration uncertainties). A detailed
investigation of this feature is beyond the scope of our paper, so we
will not consider it further.

The event mode data (all data for all PCUs on) were used to create
discrete FFTs using segments of 256-s data, resulting in power spectra
for the frequency range 1/256--256 Hz.  The power spectra were fitted
with a broken power-law component to fit the band-limited noise, one
or more Lorentzians for the broad bumps or QPOs, and a constant to
account for the Poisson level. Errors on the fit parameters were
calculated using $\Delta\chi^2$=1 and upper-limits were determined for
$\Delta\chi^2$=2.71, corresponding to 90\% confidence levels.

\section{Results}

The {\it RXTE}/PCA count rate curve (2.9--18.8 keV) is shown in
Figure~\ref{fig:pca}{\it a} and its corresponding 3.0--20.0 keV light
curve in Figure~\ref{fig:pca}{\it b} . A long term steady decreasing
trend is apparent, although during observation seven (60407-01-07-00)
the count rate and flux had slightly increased. The soft color
(Fig.~\ref{fig:pca}{\it c}) does not exhibit a similar smooth trend
but it is clear that during the third and fourth observations
(60407-01-03-00 and 60407-01-04-00), the source spectra became
considerably softer. During later observations the spectra were
significantly harder again. The hard color (Fig.~\ref{fig:pca}{\it c})
shows fewer but more erratic fluctuations, demonstrating that the
softening of the spectra during observation three and four is mainly
due to variations in the spectra at relatively low energies ($<$8
keV).

In Figure~\ref{fig:powerspectra}, several typical power spectra of SAX
J1711.6--3808 are plotted. The power spectra of the source is
dominated by a strong broad band-limited noise component, which can be
modeled using a broken power-law with a break frequency ($\nu_{\rm
break}$) of 0.2--2.0 Hz. Superimposed on this noise components, a
broad bump or QPO is present with frequencies ($\nu_{\rm QPO}$) of
1--7 Hz.  However, we note that the statistics of the data are not
very constraining and other fit functions (e.g., only a set of
Lorentzians) can be used to fit the data equally well. Also, the broad
bump or QPO might have substructure due to sub-harmonics or overtones
(note that only during observation three the first overtone could be
detected at a $>3\sigma$ level; see Tab.~\ref{tab:log}), however, the
data do not allow to constrain this further.  Only during observations
three, four, and seven, the bumps on top of the band-limited noise can
be considered true QPOs with Q values (frequency/FWHM) $>$2
(Fig.~\ref{fig:powerspectra}{\it b}). This constitutes the first
detections of QPOs for SAX J1711.6--3808.

The power spectra are very similar during the highest observed count
rates (e.g., Fig.~\ref{fig:powerspectra}{\it a}, observation
60407-01-01-00) and lowest observed ones (e.g.,
Fig.~\ref{fig:powerspectra}{\it c}, 60407-01-05-00), with a $\nu_{\rm
break}$ and $\nu_{\rm QPO}$ around 0.5 Hz and 1--2 Hz, respectively
(see Tab.~\ref{tab:log} for a detailed listing of the power spectral
results). However, when the source was at intermediate count rates
(e.g., during the third and fourth observations) the power spectra
still had the same shape, but the characteristic frequencies had
shifted to higher values (around 2 Hz and 7 Hz, respectively;
Fig.~\ref{fig:powerspectra}{\it b}). These difference is also apparent
in Figures~\ref{fig:pca}{\it d} and~\ref{fig:pca}{\it f}, which show
$\nu_{\rm break}$ and $\nu_{\rm QPO}$ as a function of time.  There is
no apparent correlation between the count rate or flux and the
properties of the broken power-law component and the QPO (i.e., their
frequencies). But the frequencies of both components are well
correlated with the soft color: the frequencies increase when the
color decreases (compare, e.g., Fig.~\ref{fig:pca}{\it c}
with~\ref{fig:pca}{\it d}). Such a correlation is not present between
the frequencies and the hard color.

There is also a good correlation between the break frequency and the
QPO frequency, which is exactly the correlation found for other LMXBs
(Wijnands \& van der Klis 1999), demonstrating that the rapid X-ray
variability in SAX J1711.6--3808 is closely related to that seen in
those other systems. For observation 60407-01-07-00, the QPO frequency
seems to be higher than expected from this correlation; however, it is
possible that only the first overtone of the QPO could be detected and
not its fundamental (see also Wijnands \& van der Klis 1999 for a
general discussion of this effect on their correlation). This would
put this observation exactly on the correlation. We searched for this
possible fundamental but fitting a Lorentzian with a fixed frequency
at half the QPO frequency, did not reveal a significant detection
(1.6$\sigma$ detection for a FWHM of $\sim$1 Hz) and a 95\% confidence
level upper limit on its strength is 6.7\% rms. This limit is not very
stringent and therefore we cannot exclude that indeed the detected QPO
is the first overtone and not the fundamental. Despite these
uncertainties, it is clear that the rapid variability of SAX
J1711.6--3808 is very similar to what has been found for other LMXBs.
Another correlation is that the strength of the band-limited noise and
the QPO tend to decrease when the frequencies of those components
increase. These correlations are not exact, but they are clearest for
the QPO strength.

\begin{figure}[]
\begin{center}
\begin{tabular}{c}
\psfig{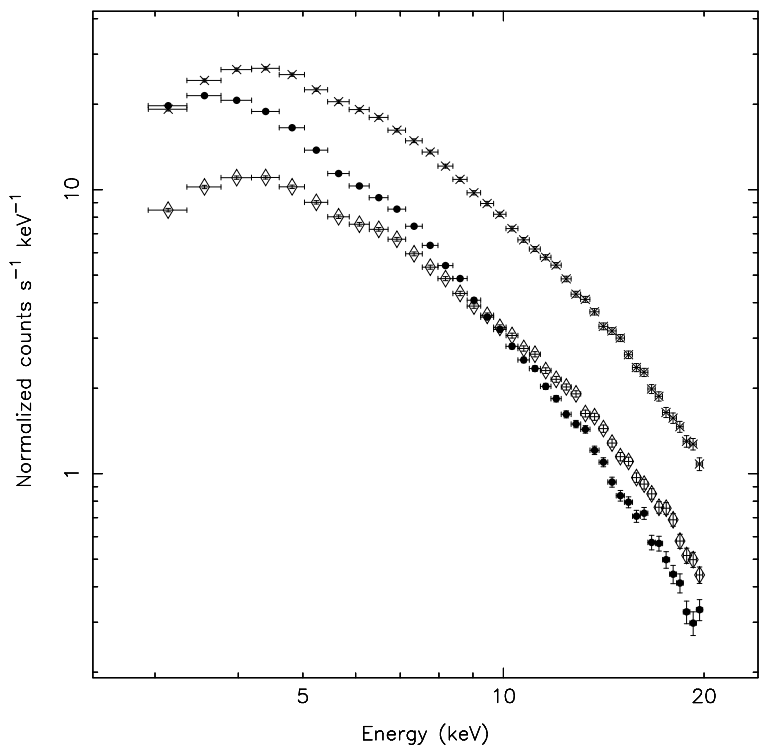}
\end{tabular}
\figcaption{Energy spectra of the observations 60407-01-01-00
(crosses), 60407-01-04-00 (bullets), and 60407-01-05-00 (diamonds).
\label{fig:spectra} }
\end{center}
\end{figure}

Typical energy spectra of SAX J1711.6--3808 are shown in
Figure~\ref{fig:spectra}. In general, the spectra can be adequately
fitted with a power-law model (but including a Gaussian component as
explained above) with a power-law index of roughly $\sim$2 (see
Tab.~\ref{tab:log} for a detailed overview of the spectral
results). Including a multi-color disk black-body component in the
fit, did not statistically improved the fit for most observations.
However, during observation three and four, an extra soft component
was necessary, which was fitted with a multi-color disk black-body
model with a characteristic temperature of $\sim$0.8 keV (see
Tab.~\ref{tab:log}). In Figure~\ref{fig:spectra}, it can be clearly
seen that observation four was considerably softer than observation
one, with almost similar fluxes at the lowest energies but
considerably lower fluxes at higher energies. The spectrum obtained
during observation five was also considerably harder than the one for
observation four, which is evident by the fact that both spectra cross
each other at around 10 keV (Fig.~\ref{fig:spectra}).  No apparent
correlation is present between the power-law index and time or the
count rate/fluxes, but a correlation is present between this parameter
and the strength of the rapid X-ray variability. Both the strength of
the broken-power law and the QPO tend to decrease when the power-law
index increases.

\section{Discussion}

We report on the spectral and timing behavior of the new X-ray
transient SAX J1711.6--3808 during the decay of its 2001 outburst. The
rapid X-ray variability of SAX J1711.6--3808 can be consistently
described by a broad band-limited noise component with a break
frequency between 0.2 and 2 Hz. On top of this noise component a broad
bump or a narrow QPO (only during observation three, four, and seven)
is detected, whose frequency is strongly correlated with the break
frequency. The frequencies of both components are not correlated with
the X-ray luminosity, which, except for observation seven, decreases
monotonically with time. The X-ray spectrum of the source is also not
correlated with the luminosity but it is correlated with the timing
properties. The spectrum is considerably softer (the inclusion of a
soft component on top of the power-law is required) when the
characteristic frequencies in the power spectra are higher
(observations three and four) compared with when they are
significantly lower (the other observations). From our results, it is
clear that during our third and fourth observations, the source was in
a different, softer, state than during the other six observations. The
occurrence of these states does not correlate with the X-ray
luminosity.  If the X-ray luminosity is a good indication of the mass
accretion rate then the mass accretion rate monotonically decreased
(except for observation seven) during the observations, but then the
mass accretion rate is decoupled from the state transitions. However,
it is also possible that still the mass accretion rate is coupled with
the different states, but then the X-ray luminosity cannot be a good
tracer of \mdot. In both scenarios it is difficult to explain why
certain parameters in SAX J1711.6--3808 are coupled to the mass
accretion rate but others are not.

In this respect, it is useful to consider that we only have the fluxes
in a limited energy range (3--20 keV), which might not be an accurate
indication for the total luminosity. A considerable amount of source
flux could be hidden at luminosities above 20 keV.  However, one
crucial point can be made in this respect. During the first two of our
observations, the source is the brightest in the 3--20 keV energy
range and has relatively hard power-law spectra. However, during
observation 3 and 4 the source is considerably dimmer and the spectra
are much softer with steeper power-law tails. Therefore, no extra flux
is expected to be present for those observations at higher energies
($>$20 keV, if extrapolating the spectral shape) which would make
these observations brighter than the first two. This strongly suggest
that the total X-ray flux during observation 3 and 4 was significantly
lower than during observation 1 and 2.  However, the spectra and the
rapid X-ray variability of the source clearly demonstrate that the
source had made a transit into a softer state, usually seen at higher
luminosities. Therefore, the conclusion seems to be warranted that the
X-ray luminosity is decoupled from the spectral and timing properties
of the source.

SAX J1711.6--3808 is not the only source for which several source
characteristics seem to be decoupled from the X-ray luminosity.  As
explained in the introduction, for the neutron star systems it is
assumed that their correlated spectral-timing behavior is a good trace
of the mass accretion rate and not the X-ray luminosity, mainly
because of their correlations with other source parameters such as the
emission at other wavelengths (optical, UV, line flux; Hasinger et
al. 1990; Vrtilek et al. 1990, 1991; van Paradijs et al. 1990, Hertz
et al. 1991; Augusteijn et al. 1992 ) and the X-ray burst properties
(e.g., van der Klis et al. 1990; Muno et al. 2000; Franco 2001; van
Straaten et al. 2001).  However, Homan et al. (2001b) has pointed out
that not all the behavior of the bright neutron-star LMXB GX~17+2 can
be explained when assuming the above picture.  Clearly, even for
neutron star systems it is not completely settled how the different
source properties depend on mass accretion rate. For BHCs, this
situation is even less well understood.  Homan et al. (2001a)
demonstrated that the state behavior of the X-ray transient and BHC
XTE~J1550--564 was highly complex and completely decoupled from the
luminosity, similar to what we observed for SAX~J1711.6--3808
(although in much lesser degree than what was observed for
XTE~J1550--564, which could be due to the fewer observations and the
lower luminosity of SAX J1711.6--3808). This demonstrates that BHCs
can exhibit the same ambiguity as for the neutron-star systems with
respect to how to trace \mdot.

At the moment, we cannot determine conclusively the nature of the
compact object in SAX J1711.6--3808. Both the neutron star and the
black hole systems have similar spectral and timing properties as we
observe for SAX J1711.6--3808. If the source harbors a neutron star
primary, then SAX J1711.6--3808 would most likely be of the atoll-type
(see Hasinger \& van der Klis 1989 for the neutron star
classification) and the observed different states can then be
identified with the island state (the hard states) and the lower
banana branch (the soft state). However, the source might harbor a
black hole.  The spectral data of observation three and four required
the inclusion of a soft component. The obtained MCD temperature of
$\sim$0.8 keV is typically what has been observed for BHCs, but still
a neutron star primary cannot be excluded. If indeed a black-hole is
present in SAX J1711.6--3808, then it would be the second BHC LMXB,
after XTE J1550--564, for which it has been shown that the state
transitions are not coupled to the luminosity of the source. The
observed states can then be identified with the low state (the hard
states) and the intermediate state (the soft state).

The main questions which remain are: (1) what causes this decoupling
between the correlated spectral and timing behavior from the X-ray
luminosity, and (2) which of the source properties traces \mdot~best?
The answer to the first question remains elusive. Moreover, both
neutron-star and black-hole systems can exhibit this decoupling and it
is even unclear if the same physical mechanism is responsible for this
in both source types.  The second question is also difficult to
answer. Homan et al. (2001a) suggested that the states in the BHC XTE
J1550--564 are decoupled from the mass accretion rate. However, if in
this source and in the neutron star systems the same process is behind
the decoupling of the states with luminosity, then it is possible that
also for XTE J1550--564 the states are still coupled to \mdot~but the
luminosity is not, similar to the neutron star systems. The answers to
the questions are especially important for the X-ray transients.  If
it can be proven that the X-ray luminosity is decoupled from \mdot,
then this would have profound implications about our understanding of
the outbursts observed for those transients. Usually, when those
outburst light curves are modeled, it is implicitly assumed that the
luminosity is a direct trace of \mdot.  The results obtained from
those models will loose their validity if this assumption turns out to
be false. It is possible that even among systems with the same type of
compact object, some light curves trace \mdot, while others do
not. This may account (partly) for the different outburst light curve
profiles observed for X-ray transients.

\begin{figure}[]
\begin{center}
\begin{tabular}{c}
\psfig{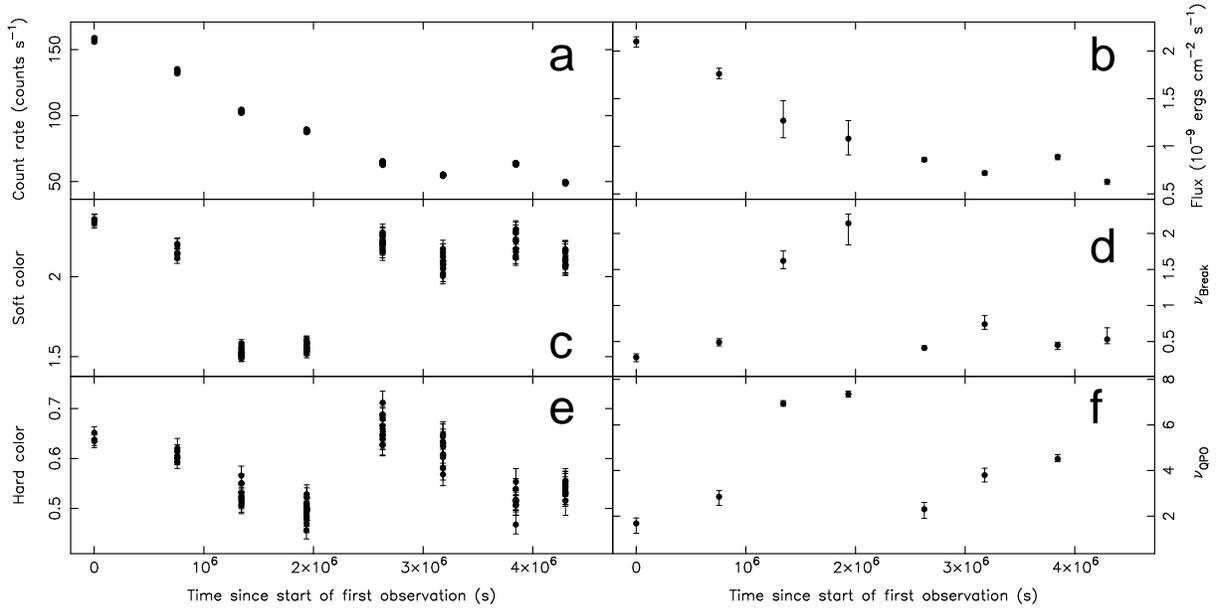}
\end{tabular}
\figcaption{The 2.9--18.8 keV PCA count rate curve ({\it a}, PCU 2
only, the count rates are background subtracted), the 3.0--20.0 keV
unabsorbed flux ({\it b}), the soft color ({\it c}, the count rate
ratio between 4.1--7.5 keV and 2.9--4.1 keV), the break-frequency
({\it d}), the hard color ({\it e}, the count rate ratio between
11.4--18.8 keV and 7.5--11.4 keV), and the QPO frequency ({\it f}) as
a function of time since the start of the first observation
(60407-01-01-00; 2001 March 18). The bin size in {\it a}, {\it c}, and
{\it e} is 256 seconds, to show that the variability in those
quantities during each observation is less than the variability
between observations.  For the other panels, the whole data through
out the separate observations was used (see Tab.~\ref{tab:log} for the
total time of each observation), to increase sensitivity.
\label{fig:pca} }
\end{center}
\end{figure}

\begin{figure}[]
\begin{center}
\begin{tabular}{c}
\psfig{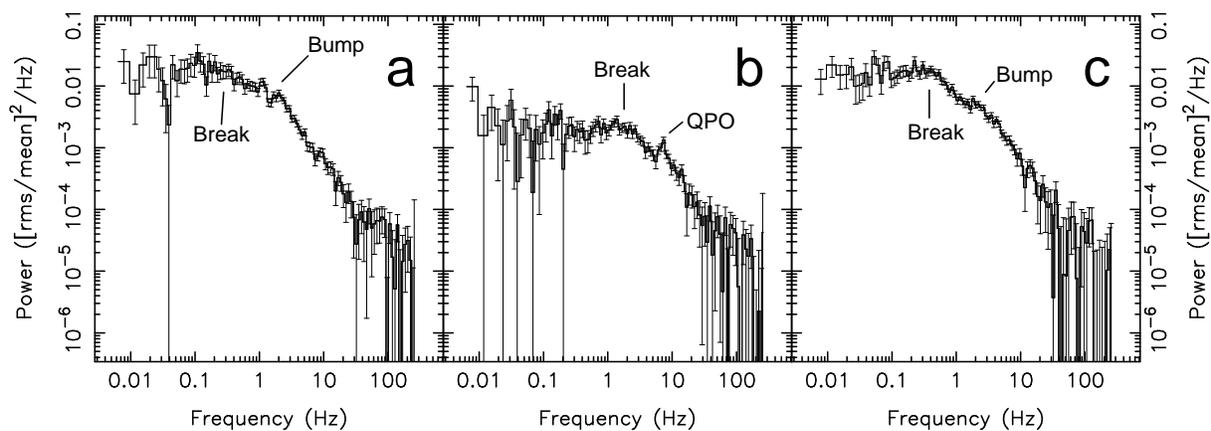}
\end{tabular}
\figcaption{Power-spectra of the observations 60407-01-01-00 ({\it
a}), 60407-01-04-00 ({\it b}), and 60407-01-05-00 ({\it c}). The
Poisson level has been subtracted.
\label{fig:powerspectra} }
\end{center}
\end{figure}

As a final point, we note that in the {\it RXTE}/PCA field of SAX
J1711.6--3808 another X-ray transient is located, the X-ray burster
SAX J1712.6--3739 (in 't Zand et al. 1999; Cocchi et al. 1999), which
is about 0.5\degr~from the position of SAX J1711.6--3808. One might
worry about any possible contamination from this source to the X-ray
flux we observe from SAX J1711.6--3808 and the possibility of this
extra flux causing the decoupling we claim. We have no information
about the X-ray state (quiescent or active) of SAX J1712.6--3739 at
the time of our observations, and on first principle we cannot exclude
any contribution of this source to the measured flux.  However, as
shown in Figure~\ref{fig:pca} {\it a} or {\it b}, during the first six
observations the source declined steadily in count rate and flux. This
behavior is expected for an X-ray transient which is decaying during
its outburst. It would be very surprising if such a trend was produced
by two independent sources, because they have to vary in X-ray
luminosity in tandem. Also, the power spectra shown in
Figure~\ref{fig:powerspectra} are what is expected when only one
source contributes to the X-ray flux. If multiple sources contributed,
then the power spectra would be distorted because of the different
timing properties of each source, unless by chance they would be very
similar.  We consider these options very unlikely and we conclude that
the flux we observe is originating from SAX J1711.6--3808, and very
likely suffers insignificant contamination from SAX J1712.6--3739.

\acknowledgments

This work was supported by NASA through Chandra Postdoctoral
Fellowship grant number PF9-10010 awarded by CXC, which is operated by
SAO for NASA under contract NAS8-39073.  This research has made use of
data obtained through the HEASARC Online Service, provided by the
NASA/GSFC and results provided by the ASM/{\it RXTE} team. We would
like to thank the referee, Keith Jahoda, for his very helpful
comments. We also would like to thank him and Jean in 't Zand for
pointing out the presence of the source SAX J1712.6--3739 in the {\it
RXTE}/PCA field of view of SAX J1711.6--3808.

\clearpage

\begin{deluxetable}{lcccccccc}
\tabletypesize{\tiny}
\tablecolumns{9}
\tablewidth{0pt} 
\tablecaption{Rapid X-ray variability and spectral parameters of each observation\label{tab:log}}
\tablehead{
Observation (60407-01-)   & 01-00                 & 02-00                  & 03-00               & 04-00               & 05-00               & 06-00               & 07-00               & 08-00}
\startdata
Obs. time (2001 UT)       & March 18              & March 26--27           & April 2             & April 9             & April 17            & April 24            & May 1               & May 6--7 \\
                          & 05:42--06:20          & 23:56--00:32           & 18:22--19:09        & 15:25--17:14        & 15:41--17:43        & 00:51--01:53        & 17:35--18:55        & 23:04--00:02 \\
Good time (ksec)          & 1.5                   & 1.4                    & 2.5                 & 3.7                 & 3.4                 & 2.6                 & 2.3                 & 2.5                  \\
PCUs on                   & All                   & 0,2,3$^a$              & 0,2,4               & All$^a$             & All                 & 0,2,3               & 0,2,4               & 0,2,3 \\
Count rate$^b$ (cts \pers)& 156                   & 134                    & 103                 & 89                  & 64                  & 55                  & 64                  & 49    \\
\hline
\hline
\multicolumn{9}{c}{Power-spectrum fit parameters$^c$}\\
\hline
Broken power-law          & & & & & & & &  \\
\ \ \ \ Rms$^d$ (\%)      & 18$^{+1}_{-2}$        & 15\pp2                 & 13\pp1              & 13.6\pp0.5          & 16.7\pp0.5          & 18$^{+1}_{-2}$      & 13\pp1              & 12\pp1 \\  
\ \ \ \ $\nu_{\rm break}$ (Hz) & 0.28\pp0.06      & 0.49\pp0.05            & 1.6\pp0.1           & 2.1$^{+0.1}_{-0.3}$ & 0.41\pp0.03         & 0.7\pp0.1           & 0.45\pp0.06         & 0.5$^{+0.2}_{-0.1}$ \\
\ \ \ \ $\Gamma_1^e$      & --0.4$^{+0.1}_{-0.2}$ & --0.4\pp0.1            & --0.29\pp0.09       & --0.04\pp0.05       & --0.10\pp0.06       & 0.02\pp0.07         & --0.6$^{+2}_{-3}$   & --1.6$^{+0.9}_{-1.7}$    \\
\ \ \ \ $\Gamma_2^e$      & 1.05\pp0.05           & 1.2$^{+0.2}_{-0.1}$    & 1.4$^{+0.3}_{-0.1}$ & 1.11\pp0.08         & 1.2                 & 1.2$^{+0.2}_{-0.1}$ & 1.2\pp0.1           & 1.0\pp0.1           \\
QPO                       & & & & & & & &  \\
\ \ \ \ Rms$^d$ (\%)      & 14$^{+3}_{-2}$        & 12$^{+3}_{-2}$         & 5.4$^{+1.2}_{-0.7}$ & 4.2\pp0.5           & 14\pp1              & 8$^{+2}_{-1}$       & 5.0$^{+1.0}_{0.8}$  & $<$7                  \\
                          &                       &                        & 6$^{+3}_{-1}$       &                     &                     &                     &                     &  \\
\ \ \ \ FWHM (Hz)         & 2.9$^{+0.6}_{-0.5}$   & 4\pp1                  & 1.4$^{+0.9}_{-0.5}$ & 1.4$^{+0.5}_{-0.3}$ & 5.6\pp0.7           & 2\pp1               & 0.8$^{+0.4}_{-0.5}$ & -- \\
                          &                       &                        & 4$^{+5}_{-2}$       &                     &                     &                     &                     &  \\
\ \ \ \ $\nu_{\rm QPO}$   & 1.7$^{+0.2}_{-0.4}$   & 2.9$^{+0.3}_{-0.4}$    & 6.9\pp0.1           & 7.4\pp0.1           & 2.3$^{+0.3}_{-0.4}$ & 3.8\pp0.3           & 4.5$^{+0.2}_{-0.1}$ & -- \\
                          &                       &                        & 13.4$^{+0.5}_{-1.3}$&                     &                     &                     &                     &  \\
$\chi^2$/dof              & 112/92                & 78/92                  & 88/89               & 82/92               & 87/92               & 81/92               & 69/92               & 94/95 \\
\hline
\hline
\multicolumn{9}{c}{Energy-spectrum fit  parameters$^f$}\\
\hline
Power-law                 & & & & & & & &  \\
\ \ \ \ Index             & 1.87\pp0.01           & 2.04\pp0.02            & 2.12\pp0.07         & 2.36\pp0.06         & 1.89\pp0.02         & 2.06\pp0.02         & 2.33\pp0.02         & 2.16\pp0.02 \\
\ \ \ \ Normalization     & 0.51\pp0.02           & 0.60\pp0.02            & 0.38\pp0.06         & 0.6\pp0.1           & 0.22\pp0.01         & 0.25\pp0.01         & 0.55\pp0.02         & 0.27\pp0.01 \\
\ \ \ \ Flux$^g$          & 1.93\pp0.06           & 1.62\pp0.05            & 0.9\pp0.1           & 0.84\pp0.14         & 0.80\pp0.02         & 0.66\pp0.02         & 0.75\pp0.03         & 0.57\pp0.03 \\
MCD black-body            & & & & & & & &  \\
\ \ \ \ T (keV)           & --                    & --                     & 0.86\pp0.04         & 0.78\pp0.06         & --                  & --                  & --                  & --\\
\ \ \ \ Normalization     & --                    & --                     & 120\pp30            & 108\pp50            & --                  & --                  & --                  & --\\
\ \ \ \ Flux$^g$          & --                    & --                     & 0.24\pp0.06         & 0.11\pp0.05         & --                  & --                  & --                  & --\\
Total flux$^g$            & 2.10\pp0.06           & 1.76\pp0.06            & 1.3\pp0.2           & 1.1\pp0.2           & 0.86\pp0.02         & 0.72\pp0.02         & 0.89\pp0.03         & 0.63\pp0.03\\
$\chi^2_{\rm red}$        & 1.69                  & 1.09                   & 0.86                & 0.96                & 1.18                & 0.73                & 1.20                & 1.02 \\  
\enddata 

\tablenotetext{a}{For 02-00, PCU 0 and 2 were always on but PCU 3 only
for 0.65 ksec. For 04-00, all detectors were on for the first 3.5
ksec, but only PCU 0 and 2 were on during the remaining 0.2 ksec.}

\tablenotetext{b}{The count rates are for PCU 2 only and for 2.9--18.8
keV. The count rates are background subtracted.}

\tablenotetext{c}{All errors are determined using $\Delta\chi^2=1$ and
the upper limits are for 95\% confidence levels}

\tablenotetext{d}{The rms amplitudes are for the full 2--60 keV energy
range of the {\it RXTE}/PCA.}

\tablenotetext{e}{$\Gamma_1$ is the slope of the power spectra for
frequencies lower than $\nu_{\rm break}$ and $\Gamma_2$ for
frequencies higher than that. $\Gamma_2$ was fixed 1.2 (the averaged
of the observations) for observation 5 because it could not be
constrained during the fit.}

\tablenotetext{f}{The column density was fixed to $1.4\times10^{22}$
cm$^{-2}$ (except for observation 7, see text). The errors are for
90\% confidence levels.}

\tablenotetext{g}{The fluxes are for 3.0--20.0 keV and in units of
$10^{-9}$ \funit. The power-law and MCD fluxes are absorbed but the
total flux is unabsorbed.}

\end{deluxetable}


\begin{references}

\reference{}Arnaud, K. \& Dorman, B. XSPEC v. 11.0

\reference{}Augusteijn, T., et al. 1992,
\aap, 265, 177

\reference{}Christian, D. J. \& Swank, J. H. 1997, \apjs, 109, 177

\reference{}Cocchi, M., Natalucci, L., in 't Zand, J., Heise, J.,
Muller, J. M., Celidonio, G., Di Ciolo, L. 1999, \iaucirc, 7247

\reference{}Cui, W., Shrader, C. R., Haswell, C. A., Hynes,
R. I. 2000, \apj, 535, L123
 
\reference{}Dickey, J. M. \& Lockman, F. J. 1990 \araa, 28, 215

\reference{}Dieters, S. W., Belloni, T., Kuulkers, E., Woods, P., Cui,
W., Zhang, S. N., Chen, W., van der Klis, M., van Paradijs, J., Swank,
J., Lewin, W. H. G., Kouveliotou, C. 2000, \apj, 538, 307

\reference{}Franco, L. M. 2001, ApJ in press (astro-ph/0009189)

\reference{}Hasinger, G. \& van der Klis, M., 1989, \aap, 225, 79

\reference{}Hasinger, G., van der Klis, M., Ebisawa, K., Dotani, T.,
Mitsuda, K. 1990, \aap, 235, 131

\reference{}Hertz, P., Vaughan, B., Wood, K. S., Norris, J. P.,
Mitsuda, K., Michelson, P. F., Dotanti, T. 1992, \apj, 396, 201

\reference{}Homan, J., Wijnands, R., \& van der Klis, M. 2000,
\iaucirc, 7121

\reference{}Homan, J., Wijnands, R., van der Klis, M., Belloni, T., van
Paradijs, J., Klein-Wolt, M., Fender, R., M\'endez, M. 2001a, \apjs 132,
377

\reference{}Homan, J., van der Klis, M., Jonker, P. G., Wijnands, R.,
Kuulkers, E., M\'endez, M., Lewin, W. H. G. 2001b, ApJ submitted
(astro-ph/0104323)

\reference{}in 't Zand, J., Heise, J., Bazzano, A., Cocchi, M., Smith,
M. J. S. 1999, \iaucirc, 7243

\reference{}in 't Zand, J. J. M., Kaptein, R. G., \& Heise, J. 2001,
\iaucirc, 7582

\reference{}M\'endez, M. \& van der Klis, M. 1997, \apj, 479, 926

\reference{}M\'endez, M., Belloni, T., \& van der Klis, M. 1998, \apj,
499, L187

\reference{}Miller, J. M., Wijnands, R., Homan, J., Belloni, T.,
Pooley, D., Corbel, S., Kouveliotou, C., van der Klis, M., Lewin,
W. H. G. 2001, \apj letters, submitted (astro-ph/0105371)

\reference{}Muno, M. P., Fox, D. W., Morgan, E. H., Bildsten, L. 2000,
\apj, 542, 1016

\reference{}Remillard, R. A., McClintock, J. E., Sobczak, G. J., Bailyn,
C. D., Orosz, J. A., Morgan, E. H., \& Levine, A. M.  1999a, \apj, 517,
L127
  
\reference{}Remillard, R. A., Morgan, E. H., McClintock, J. E., Bailyn,
C. D., Orosz, J. A. 1999b, \apj, 522, 397


\reference{}Sobczak, G. J., McClintock, J. E., Remillard, R. A., Cui,
W., Levine, A. M., Morgan, E. H., Orosz, J. A., Bailyn, C. D. 2000,
\apj, 531, 537

\reference{}Strohmayer, T. E. 2001, \apj, 522, L49

\reference{}Tanaka. Y, \& Lewin, W. H. G.  1995, In: {\it X-ray
Binaries}, W. H. G. Lewin, J. van Paradijs, \& E. P. J. van den Heuvel
(eds.), Cambridge University Press, p. 126


\reference{}van der Klis, M.  1995, In: {\it X-ray Binaries},
W. H. G. Lewin, J. van Paradijs, \& E. P. J. van den Heuvel (eds.),
Cambridge University Press, p. 252
 
\reference{}van der Klis, M. 2000, \araa, 38, 717

\reference{}van der Klis, M., Hasinger, G., Damen, E., Pennix, W., van
Paradijs, J., Lewin, W. H. G. 1990, \apj, 360, L19

\reference{}van Paradijs, J., Allington-Smith, J., Callanan, P.,
Charles, P. A., Hassall, B. J. M., Machin, G., Mason, K. O., Naylor,
T., Smale, A. P. 1990, \aap, 235, 156

\reference{}van Straaten, S., van der Klis, M., Kuulkers, E.,
M\'endez, M. 2001, \apj, 551, 907

\reference{}Vrtilek, S. D., Raymond, J. C., Garcia, M. R., Verbunt,
F., Hasinger, G., Kurster, M. 1990, \aap, 235, 162

\reference{}Vrtilek, S. D., Pennix, W., Raymond, J. C., Verbunt, F.,
Hertz, P., Wood, K., Lewin, W. H. G., Mitsuda, K. 1991, \apj, 376, 278

\reference{}Wijnands, R. \& van der Klis, M. 1999, \apj, 514, 939

\end{references}
\end{document}